\documentclass{aa}
\usepackage{graphicx}
\usepackage{epsfig}
\usepackage{graphics}
\begin{document}%
  \title{High-energy emission from jet-clump interactions in microquasars}
%\subtitle{}

   \author{A.~T. Araudo\inst{1,2,}\thanks{Fellow of CONICET, Argentina},
           V. Bosch-Ramon\inst{3} \and
           G.~E. Romero\inst{1,2,}\thanks{Member of CONICET, Argentina}
           %\and V. Bosch-Ramon\inst{3}%\thanks{}
          }

   \offprints{Anabella T. Araudo: \\ {\em aaraudo@fcaglp.unlp.edu.ar}}
   \titlerunning{High-energy emission from jet-clump interaction}

\authorrunning{A.T. Araudo et al.}  
\institute{
Instituto Argentino de Radioastronom{\'\i}a (CCT La Plata, CONICET), 
C.C.5, 1894 Villa Elisa,  Buenos Aires, Argentina
\and Facultad de Ciencias Astron\'omicas y Geof\'{\i}sicas,
Universidad Nacional de La Plata, Paseo del Bosque, 1900 La Plata,
Argentina 
\and Max Planck Institut f\"ur Kernphysik, Saupfercheckweg
1, Heidelberg 69117, Germany}

\date{Received / Accepted}

% \abstract{}{}{}{}{}
% 5 {} token are mandatory

\abstract 
{High-mass microquasars are binary systems consisting of a 
massive star and an accreting compact object from which relativistic jets are
launched.   
There is considerable observational evidence  
that winds of massive stars are clumpy. Individual clumps may interact with
the jets in high-mass microquasars to produce outbursts of high-energy 
emission. 
Gamma-ray flares have been detected in some high-mass X-ray
binaries, such as Cygnus X-1, and probably in LS 5039 and LS I+61 303.}
{We predict the high-energy emission produced by the interaction 
between a jet and a clump of the stellar wind in a high-mass microquasar.}
{Assuming a hydrodynamic scenario for the jet-clump interaction,
we calculate the spectral energy distributions produced by
the dominant non-thermal processes: relativistic bremsstrahlung, synchrotron 
and inverse Compton  radiation, for leptons, and for hadrons, proton-proton 
collisions.}  
{Significant levels of emission in X-rays (synchrotron),
high-energy gamma rays (inverse Compton), and very high-energy gamma rays 
(from the decay of neutral pions) are predicted, with luminosities in
the different domains in the range $\sim 10^{32}$-$10^{35}$~erg~s$^{-1}$.
The spectral energy distributions vary strongly depending on 
the specific conditions.}
{Jet-clump interactions may be detectable at high and very high energies, 
and provide an explanation for the fast TeV variability
found in some high-mass X-ray binary systems.
 Our model can help to infer
information about the properties of jets and clumpy winds by means of
high-sensitivity gamma-ray astronomy.}

\keywords{Gamma-rays: theory -- X-rays: binaries -- Radiation mechanisms:
non-thermal}

\maketitle
%______________
\section{Introduction}

The mass loss in massive and hot stars is
understood to occur by means of supersonic inhomogeneous winds structured as
very dense and small clumps embedded in large regions of tenuous plasma.
This idea is supported by considerable observational evidence 
of the clumpy structure of these winds (e.g. Puls et al. 2006, Owocki
\& Cohen 2006, Moffat 2008). 
However, as a consequence of the  high spatial resolution necessary for a 
straightforward  
detection of the clumps, all the evidence is indirect. Hence, 
properties of clumps, such as size, density, and number, are not well-known.  

Some massive stars are accompanied by a compact object,
to which they transfer matter. This process leads to
the formation of an accretion disk around the compact object and,
in high-mass microquasars (HMMQs), to the generation of bipolar 
relativistic outflows (e.g. Mirabel \& Rodr{\'\i}guez 1999).

Non-thermal emission has been observed in microquasar jets from radio 
(e.g. Rib\'o 2005) to X-rays (e.g. Corbel et al. 2002). 
At higher energies, gamma-ray radiation could also be produced in jets
(e.g. Bosch-Ramon et al. 2006).
A TeV flare was detected by MAGIC from the 
HMMQ Cygnus~X-1 (Albert et al. 2007). 
Transient gamma-ray events may also have
been detected from the high-mass X-ray binaries   
LS~5039 and LS~I+61~303
by HESS (Aharonian et al. 2005) and MAGIC (Albert et al. 2006), respectively, 
as suggested by Paredes (2008). 
In addition, the HMMQ Cygnus X-3 might have been observed flaring in
the GeV range by \emph{AGILE} (ATels 1492, 1547 and 1585, see however Atel 
1850). Both this instrument and \emph{Fermi} also found several transient
GeV sources in the Galactic plane without known counterpart (ATel 1394,
Abdo et al. 2009).
All this strongly variable gamma-ray emission may have a similar origin.
For instance, the interaction between the jet and the
stellar wind of the companion star could produce 
TeV flares (e.g., Romero et al. 2003, Romero \& Orellana 2005, Romero et al. 
2007, Albert et al. 2007, Perucho \& Bosch-Ramon 2008, Owocki et al. 2009). 

In this paper, we propose a model to explain these 
gamma-ray flares, based on the interaction between 
the jets of a 
HMMQ with wind inhomogeneities. 
The clumps can eventually penetrate the jet, 
leading to transient non-thermal activity that may
release a significant fraction of the jet kinetic luminosity in the
form of synchrotron, inverse Compton (IC), and proton-proton ($pp$) 
$\pi^0$-decay emission. 

This work is organized as follows. In the next section, we describe the 
main characteristics of the scenario adopted and estimate the relevant 
timescales of the jet-clump interaction;
in Sect. 3, we study the acceleration of particles and estimate the
non-thermal radiative losses; 
in Sect. 4, we describe the calculation of the non-thermal emission and 
present the main results; and finally, in Sect. 5, we draw 
some conclusions. CGS units are used consistently throughout the paper.

\section{The physical scenario}
\label{scenario}

\begin{figure}
\includegraphics[angle=0, width=0.45\textwidth]{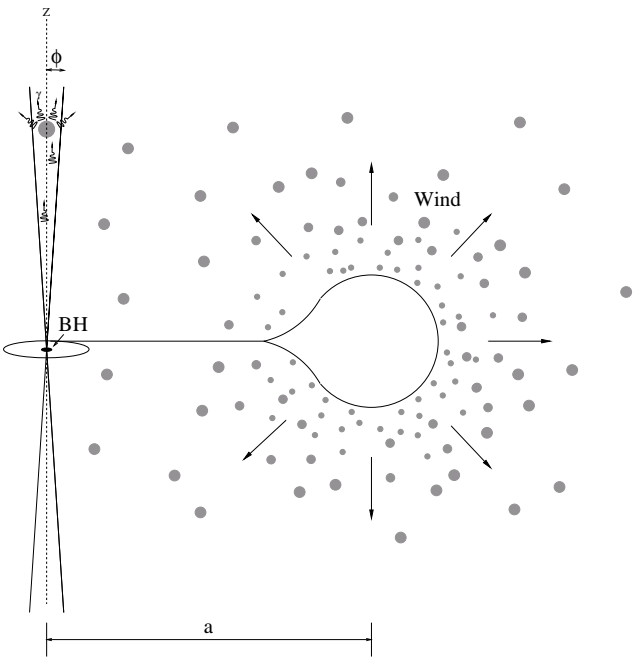}
\caption{Sketch of a HMMQ with clumpy stellar wind, adapted from Romero 
et al. (2007).}\label{fig_1}
\end{figure}

To study the interaction between a clump of the stellar wind 
and a jet in a HMMQ, we adopt a scenario
with similar characteristics to the binary system Cygnus~X-1. 
We fix the separation between the
compact object and the massive star to be
$a=3\times10^{12}$~cm ($0.2$ UA).
For the star luminosity and temperature, we assume that
$L_{\star}=10^{39}$~erg~s$^{-1}$ and $T_{\star} = 3\times10^4$~K,
and the
stellar mass loss rate is adopted to be $\dot{M}_{\star}=3\times
10^{-6}\,M_{\odot}$~yr$^{-1}$, with a terminal wind velocity  
$v_{\rm w} \sim 2.5\times10^8$~cm~s$^{-1}$.
A sketch of the scenario is presented in Fig. \ref{fig_1}.

\subsection{Clump model}

The clump is assumed to be spherical and homogeneous. 
Given the uncertainties in the clump parameters,
we consider two values of its size: $R_{\rm c}= 10^{10}$ and $10^{11}$~cm 
(i.e., $\sim 3\times(10^{-3}-10^{-2})\,a$). 
For effective jet penetration, a large density contrast between the clump
and the jet is required. Therefore, we assume dense clumps with 
$n_{\rm c} = 10^{12}$~cm$^{-3}$, which correspond to a porous clumpy wind 
with a filling factor (clump versus interclump volume ratio)  
of $f=\dot{M}_{\star}/4\pi a^2 m_p v_{\rm c} n_{\rm c} = 0.005$, 
where $m_p$ is the mass of the proton\footnote{See Owocki \& Cohen (2006) 
for the concept of a porous wind.}.
We assume that the velocity of the clumps  equals  the velocity 
of the wind, i.e., $v_{\rm c} = 2.5\times 10^8$~cm~s$^{-1}$. The temperature
of the clumps is taken as $T_{\rm c} = 10^4$~K (Krti$\check{\rm c}$ka 
\& Kub\'at 2001), which is
moderately lower than the temperature at the surface of the massive star.

\subsection{Jet model}

Radio observations demonstrated that the jets of MQs are strongly 
collimated (Miller-Jones, Fender \& Nakar 2006). We 
assume  that the jet radius is one tenth of the jet 
height, i.e. $R_{\rm j}(z) = 0.1\,z$, which corresponds to a semi-opening 
angle of $\phi = 6^{\circ}$.
The expansion velocity results in $v_{\rm exp} = 0.1\;v_{\rm j}$, 
where $v_{\rm j}$ is the velocity of the jet. This expansion velocity
for a free-expanding hidrodynamical jet implies 
a jet base Mach number $\sim v_{\rm j}/v_{\rm exp} \sim 10$, i.e. a
strongly supersonic outflow.
We consider a jet dynamically dominated by cold protons
with a weakly relativistic bulk velocity $v_{\rm j} = 0.3\,c$, i.e. a
Lorentz factor $\gamma_{\rm j} = 1.06$.
We neglect the curvature of the jet produced by its
interaction with the stellar wind. 
In HMMQ jets of luminosity $> 10^{36}$~erg~s$^{-1}$, the
geometry should not be strongly modified (Perucho \& Bosch-Ramon 2008),  
although this effect could be important in systems such as Herbig-Haro jets 
interacting with a stellar wind, as studied by Raga et al. (2009).

The kinetic luminosity of the jet is taken to
be $L_{\rm kin} \sim 3\times10^{36}\;\rm{erg\;s^{-1}}$, similar
to that of Cygnus~X-1 (e.g., Gallo et al. 2005, Russell et al. 2007). 
Using the equation

\begin{equation} 
L_{\rm kin} = \sigma_{\rm j} (\gamma_{\rm j} - 1) m_p\;c^2\,n_{\rm j}\;v_{\rm j},
\end{equation} 
where $\sigma_{\rm j}=\pi R_{\rm j}^2$ is the cross-section of the jet, we
can estimate the density $n_{\rm j}$ of the jet material in 
the laboratory reference frame (LRF).
At the height of the jet-clump interaction, which is taken to occur at 
$z=a/2$, 
we obtain $n_{\rm j} = 4.7\times10^7$~cm$^{-3}$. Thus, the ratio of the
clump to the jet densities is $\chi=2.1\times10^4$. This parameter
will be very relevant to the jet-clump interaction estimates.
The parameter values of the clump and the jet are shown in Table
\ref{parameters}.
We assume an ideal gas equation of state, $P=nkT$, for both the clump 
and the jet, where $n$ is the gas density.

\begin{table}[]
\begin{center}
\caption{Adopted parameters in this work.}\label{parameters}
\begin{tabular}{lcc}
\hline 
Parameter description [units] & Clump & Jet \\
\hline 
Radius [cm] & $10^{10}-10^{11}$ & $1.5\times10^{11}$ \\  
Velocity [$\rm cm\;s^{-1}$] & $2.5\times10^8$ & $10^{10}$ \\  
Density [$\rm cm^{-3}$] & $10^{12}$ & $4.7\times10^7$ \\  
\hline
{} & Binary system & {} \\  
\hline  
System size [cm] & $3\times10^{12}$ &{}\\
Star luminosity [erg~s$^{-1}$] & $10^{39}$ & {}\\
Star temperature [K] & $3\times10^4$ & {}\\
Mass loss rate [M$_{\odot}$ yr$^{-1}$] & $3\times10^{-6}$ &{}\\
Wind velocity [$\rm cm\;s^{-1}$] & $2.5\times10^8$ & {} \\
\hline
\end{tabular}
\end{center}
\end{table}

\subsection{Dynamics of the interaction}

Clumps are formed in the wind acceleration region 
within one stellar radius of the surface of the star 
(e.g. Puls et al. 2006) and 
some of them reach the jet/wind interface.
The very large inertia of the clumps, linked to the exceptionally
high density contrast $\chi$, allows them to cross the
boundary of the jet and fully penetrate into it. 
A large value of  $\chi$ is also required to ensure that
the clump is not strongly affected when penetrating into the jet. 
 
To study the physical processes of the interaction, 
we consider the collision of  a single clump with the jet.
For simplicity, we assume that, on the relevant spatial scales of
the interaction, the jet is cylindrical. 
In the context of this work, thermal conduction, clump expansion, magnetic
fields and gravitational forces are not dynamically relevant and will be
neglected. 

In the LRF, the clump will take a time $t_{\rm c}$, or jet penetration time, 
to fully enter the jet. 
This provides the first relevant timescale

\begin{equation} 
t_{\rm c}\sim 2\,R_{\rm c}/v_{\rm c},
\end{equation}
which is $t_{\rm c}\sim 80$ and $\sim 800$~s,
for $R_{\rm c} = 10^{10}$ and $10^{11}$~cm, respectively.
In addition, the clump crosses
the jet roughly at the wind velocity in the jet-crossing time

\begin{equation} 
t_{\rm j}\sim \frac{2\; R_{\rm j}}{v_{\rm c}} = 1.2\times10^3~\rm s.
\end{equation} 
From the moment when the clump interacts with the jet, 
the ram pressure exerted by the latter produces a shock
in the clump, which propagates in the direction of the jet
motion. Assuming that a significant fraction of the momentum flux is
transferred to the clump, the clump-crossing time, which is
the characteristic timescale of the jet-clump interaction,
can be defined to be
\begin{equation} 
t_{\rm cc}\sim \frac{2\; R_{\rm c}}{v_{\rm cs}}\sim \frac{2\; R_{\rm
c}\sqrt{\chi}}{v_{\rm j}},
\end{equation} 
where
\begin{equation} 
v_{\rm cs}\sim \frac{v_{\rm j}}{\sqrt{\chi}}
\end{equation} 
is the velocity of the shock moving through the clump, derived by 
equating the jet/clump ram pressures.
The timescale $t_{\rm cc}$ is $\sim 3\times10^2$ and $3\times 10^3$~s
for $R_{\rm c} = 10^{10}$ and $10^{11}$~cm, respectively.

\begin{figure}
\resizebox{\hsize}{!}{\includegraphics{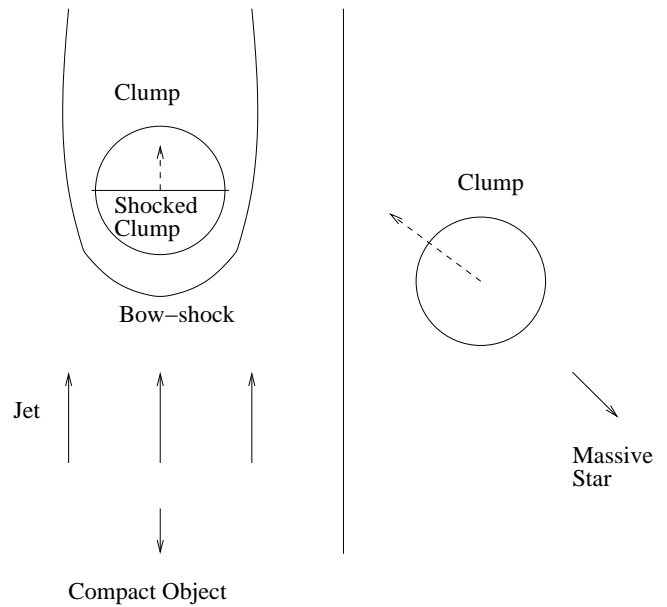}}
\caption{Sketch of the jet-clump interaction.}\label{fig_2}
\end{figure}

A shock (the bow shock) is also formed in the jet when its material 
collides with the clump. 
We assume that the bow-shock region width (or clump/bow-shock separation
distance) is $x\sim 0.2 R_{\rm c}$ (van Dyke \& Gordon 1959), 
and thus the time 
required  to reach the steady state regime is
\begin{equation} 
t_{\rm bs}\sim\frac{0.2 R_{\rm c}}{v_{\rm j~ps}},
\end{equation}  
where $v_{\rm j~ps}\sim v_{\rm j}/4$ is the jet postshock velocity in
the LRF. Therefore, $t_{\rm bs}\sim
(5/2)t_{\rm cc}/\sqrt{\chi}$, i.e. $t_{\rm bs} \ll t_{\rm cc}$. 
In Fig. \ref{fig_2},  a sketch of the situation is shown.

Once the clump is inside the jet,
the latter transfers momentum to the clump,
accelerating it to the background velocity, $v_{\rm j}$. The
acceleration can be obtained from the force exerted by the jet ram
pressure and the clump mass (e.g., Fragile et al. 2004):
\begin{equation} 
g\sim\frac{v_{\rm j}^2}{\chi R_{\rm c}}\,.
\end{equation}
The jet accelerates the clump to the velocity $v_{\rm j}$ in a
time
\begin{equation} 
t_{\rm g}\sim\frac{v_{\rm j}}{g} \sim \sqrt\chi\; t_{\rm cc},
\end{equation}
which is much longer than $t_{\rm cc}$.
When one fluid exerts a force against another fluid of different density,
the hydrodynamical Rayleigh-Taylor (RT) instability eventually develops,
leading to the perturbation and potential disruption of the clump.  The timescale
for this instability is
\begin{equation} 
t_{\rm RT}\sim \sqrt{l/g},
\end{equation}   
where $l$ is the instability length in the perturbed region.

After the bow shock is formed, the jet material 
surrounds the clump and
we have two fluids that have a large relative velocity. This situation leads to
Kelvin-Helmholtz (KH) instabilities. The timescale for a KH instability 
to develop is
\begin{equation} 
t_{\rm KH}\sim\frac{l\;\sqrt{\chi}}{v_{\rm rel}},
\end{equation}  
where $v_{\rm rel}$ is the relative
velocity between the jet material and the clump material (we assume 
$v_{\rm rel} \sim v_{\rm j}$). In addition, taking into account that
we are interested in instability
lengths similar to  the clump size, i.e. $l \sim R_{\rm c}$, we find
that $t_{\rm RT}\sim t_{\rm KH} \sim t_{\rm cc}$.  

According to the timescales estimated in the previous paragraphs, 
the clump can fully enter into the jet if 
$t_{\rm c}\approx t_{\rm cc}/5$. 
At this stage, we do not consider the penetration of the clump into the jet,
and assume that the former is completely inside the latter (i.e., the system 
has cylindrical symmetry).   
The bow shock is
formed in a time much shorter than $t_{\rm c}$ and $t_{\rm cc}$
($t_{\rm bs} \sim t_{\rm cc}/\sqrt\chi$).  
We note that the clump  might not escape the jet if
$t_{\rm j} > t_{\rm RT/KH}$.
In such a case, the clump will be destroyed inside the
jet. However, numerical simulations show that  
the instability timescales are longer than the 
clump crossing time by a factor of a few (e.g., Klein, McKee \& 
Colella 1994), i.e., $t_{\rm RT/KH} > t_{\rm cc}$.

Regarding the shock properties and given the 
particular characteristics of our scenario, the shock in the clump 
is strong, radiative, and slow, whereas the
bow shock is strong as well, but adiabatic and fast. 
For these reasons, the shocked and heated material of the clump will radiate
a non-negligible part of the energy tranferred by the jet.

\subsubsection{Thermal emission from the clump}
\label{thermal}

To estimate the  density, $n_{\rm ps}(\zeta)$,
and temperature, $T_{\rm ps}(\zeta)$, of the shocked clump at a distance 
$\zeta$ from the shock, we assume pressure equilibrium between the clump and
the jet shocked plasma and use the equation of energy flux conservation,
taking into account the main channels of thermal radiative losses:
$\Lambda(T) = 7\times10^{-19}\, T^{-0.6}$~erg~cm$^{-3}$~s$^{-1}$.
The timescale of thermal cooling equals
\begin{equation} 
t_{\rm th}= 3\times10^2\frac{T_{\rm ps}(\zeta)^{1.6}}{n_{\rm ps}(\zeta)}\,
{\rm s}\,.
\label{T}
\end{equation}   
The temperature and density in the adiabatic zone of the 
post-shock region are $8.5\times10^{6}$~K and $4\times10^{12}$~cm$^{-3}$, 
respectively. 
From these values, the shocked clump material cools in a time  
$t_{\rm th}\sim 10$~s, which is shorter than $t_{\rm cc}$, which means
a distance $\zeta_{\rm th} \sim t_{\rm th}\,v_{\rm cs}/4 \sim
2\times10^{8}$~cm from the shock front. Since $\zeta_{\rm th} <
2R_{\rm c}$, the shock in the clump is radiative. 
At distances greater than $\zeta_{\rm th}$, the density of the 
shocked matter can reach values as high as $10^{14}$~cm$^{-3}$.

Although we are mainly interested in the high-energy emission produced by the 
jet-clump interaction, we estimate for comparison the free-free emission 
generated by the shocked and heated clump.
Considering $T_{\rm ps}(\zeta)$ and $n_{\rm ps}(\zeta)$, we 
estimate the free-free luminosity by integrating the emissivity
(Lang 1999) along the shocked clump, as demonstrated in Zhekov \& Palla (2007).
The corresponding thermal luminosities are $L_{\rm th} \sim 5\times10^{30}$ and 
$5\times10^{32}$~erg~s$^{-1}$ for $R_{\rm c}=10^{10}$ and $10^{11}$~cm,
respectively, peaking in the soft X-rays. In Fig.~\ref{SEDs}, the thermal 
radiation of the  clump 
is shown together with the non-thermal emission. 

In contrast to what occurs in the clump, the bow shock
is adiabatic and fast. For this reason, it is a propitious place 
to accelerate particles to relativistic energies.

\section{Particle acceleration and non-thermal cooling}

\subsection{Particle acceleration}

In the presence of a shock and magnetic field, 
non-relativistic diffusive (Fermi I) shock acceleration can 
occur (e.g. Drury 1983). 
In the linear limit of this theory and considering Bohm diffusion, 
electrons and protons will be accelerated 
to an energy $E_{e,p}$ (where $e$ stands for electrons and $p$ 
for protons) in a time 
\begin{equation} 
t_{\rm acc} =  \eta\frac{E_{e,p}}{q\,B\,c}\,,
\label{t_acc}
\end{equation} 
where $\eta \sim (8/3) (c/v_{\rm s})^2$ for perpendicular shocks 
(Protheroe 1999). In Eq. (\ref{t_acc}), $v_{\rm s}$ is the shock 
velocity (taken to be equal to $v_{\rm j}$), $B$ is the magnetic field in the
acceleration region, $c$ is the speed of light, and 
$q$ is the electron charge.
Thus, to obtain efficient acceleration of particles, i.e., 
a short $t_{\rm acc}$,
a high value of $B$ and a strong and fast shock are necessary.

To study the jet-clump interaction, we consider two
values of the magnetic field in the jet shocked (bow-shock) region, 
$B_{\rm bs}$.
First, we consider a value of the magnetic field obtained by assuming
that the magnetic energy density is $10\,\%$
of the plasma internal energy density downstream. We adopt a sub-equipartition 
value to ensure that $B_{\rm bs}$ is dynamically negligible with respect 
to matter. Fixing
\begin{equation} 
\frac{B_{\rm bs}^2}{8\pi} = 0.1 u_p\,,
\end{equation} 
where $u_p$ is the energy density of the shocked jet gas, given by
\begin{equation} 
u_p = \frac{3}{2} P = \frac{9}{8}\;n_{\rm j}\; m_p v_{\rm j}^2\,,
\end{equation} 
we obtain  $B_{\rm bs} = 150$~G. This magnetic field, plus 
$v_{\rm s} = v_{\rm j}$, yield $t_{\rm acc}\sim 10^{-2}E_{e,p}$~s.

In addition, we adopt $B_{\rm bs} = 1$~G  to check the impact on
our results of a magnetic field significantly weaker than in the 
sub-equipartition case. 

\subsection{Non-thermal cooling processes}
\label{cooling}

Relativistic leptons lose their
energy  by different non-thermal radiative processes, such as synchrotron
radiation, IC scattering and relativistic bremsstrahlung. 
On the other hand, relativistic protons can also lose energy by means of $pp$ 
interactions. Finally, the shocked plasma can produce
thermal radiation 
if the density is high enough, as shown in Sect. \ref{thermal}.

Relativistic electrons that lose their
energy by synchrotron radiation have the following cooling time
(e.g. Ginzburg \& Syrovatskii, 1964)
\begin{equation} 
t_{\rm syn} =  \frac{4.1\times10^2}{B^2\,E_e}~\rm{s}\,.
\end{equation} 
Synchrotron 
radiation is the most efficient radiative process in the bow-shock
region in the case of $B_{\rm bs}= 150$~G, where  
$t_{\rm synch} \sim 2\times10^{-2}\,E_e^{-1}$~s. 

At the interaction height considered in this work, 
$z_{\rm int} = a/2 = 1.5\times10^{12}$~cm, the energy density of 
the photons from the star is 
$u_{\rm ph} = 2.4\times10^{2}$~erg cm$^{-3}$, the typical photon energy being
$\epsilon_0 \sim 10$~eV. For $y = \epsilon_0 E_e/(5.1\times10^5\, 
\rm{eV})^2 > 1$, i.e. $E_e > 2.6\times10^{10}$~eV,
the IC interaction occurs in the Klein-Nishina (KN) regime. 
A formula for the IC cooling time valid in both a Thompson (Th) and
KN regime in a photon field with a narrow energy distribution 
is e.g. Bosch-Ramon \& Khangulyan (2009)
\begin{equation} 
t_{\rm IC} =  \frac{6.1\times10^{12}\,\epsilon_0}{u_{\rm ph}}
\frac{(1 + 8.3y)}{\ln(1+0.2\;y)}\frac{(1 + 1.3y^2)}{(1 + 0.5y + 1.3y^2)}
~\rm{s}\,.
\end{equation} 
For $B\sim 1$ G, IC is the dominant loss leptonic channel. 
 
According to the low particle density of the shocked jet, 
$n_{\rm bs}= 4\,n_{\rm j} = 2\times10^8$~cm$^{-3}$, relativistic bremsstrahlung 
losses are negligible in the bow-shock region, the 
cooling time for a completely ionized medium being (e.g. Blumenthal \& Gould, 
1970):
\begin{equation} 
t_{\rm Brem} =  \frac{1.4\times10^{16}}{n\,Z^2\left(\ln\left(\frac{E_e}
{m_ec^2}\right) + 0.36\right)}~\rm{s}\,,
\end{equation} 
where $Z$ is the atomic number, and in our case $Z=1$.

Regarding hadronic emission, $\gamma$-rays are produced if 
relativistic protons interact with nuclei through inelastic   
collisions. 
As in the case of relativistic bremsstrahlung, the proton 
cooling time due to
$pp$ interactions depends on the density $n$ (e.g. Aharonian \& Atoyan, 
1996) related
\begin{equation} 
t_{pp} \sim  \frac{2\times10^{15}}{n}~\rm{s}\,.
\end{equation}
In the bow-shock region, this process is negligible. Otherwise, if
relativistic protons accelerated in the bow shock penetrate into the clump,
$pp$ interactions can become an efficient process to generate
gamma-rays.

\subsection{Maximum energies}

Taking into account energy gains and losses, the maximum energy achieved by
particles accelerated in the bow shock can be easily estimated. Concerning
energy losses, we consider radiative 
cooling (described above) and the escape of particles. The latter 
takes into account the advection of relativistic particles by downstream
bow-shock material ($t_{\rm adv} \sim R_{\rm c}/v_{\rm j\,ps}$)
and the diffusion of particles 
($t_{\rm diff} \sim x^2/2 D_{\rm B}$, where $x = 0.2\,R_{\rm c}$ and 
$D_{\rm B} = E_{e,p} c/3 q B_{\rm bs}$ 
is the diffusion coefficient, which is assumed to be the B$\rm{\ddot{o}}$hm 
one). The corresponding timescale is

\begin{equation} 
\tau_{\rm esc} = \rm{min}\{t_{\rm diff}, t_{\rm adv}\}\,.
\end{equation}
Given the small thickness of the bow-shock region, we assume that particles
that escape via diffusion go to the clump.

\begin{figure}
\includegraphics[angle=0, width=0.45\textwidth]{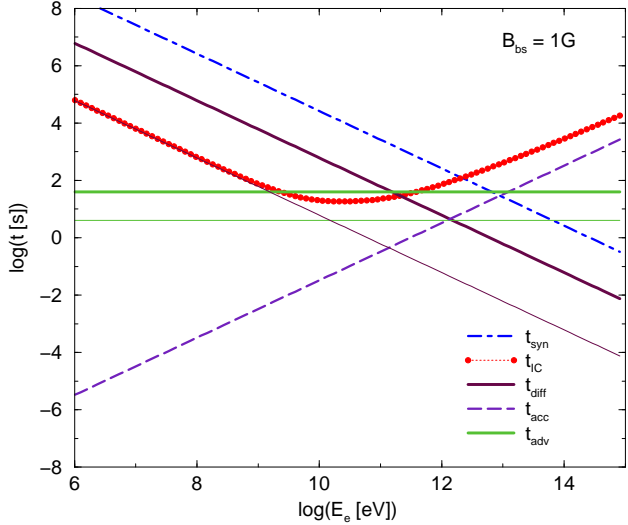}
\caption{Acceleration and radiative loss (synchrotron and IC) time for 
electrons in the bow-shock region. The advection and diffusion times are 
shown for $R_{\rm c} = 10^{10}$ (thin line) and $10^{11}$~cm
(thick line). This figure corresponds to
the case $B_{\rm bs} = 1$~G.}\label{fig_loss_1} 
\end{figure}

\begin{figure}
\includegraphics[angle=0, width=0.45\textwidth]{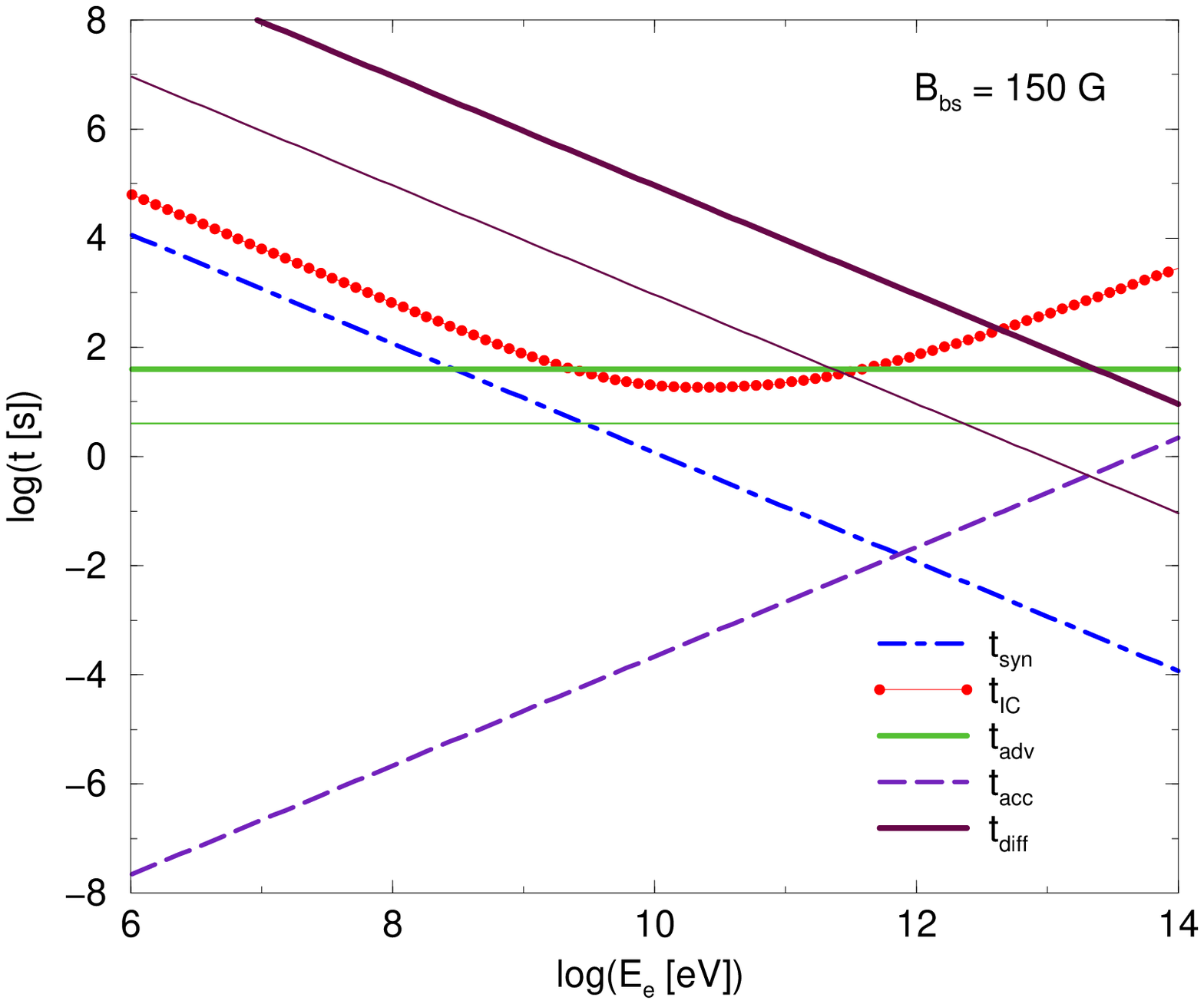}
\caption{Acceleration and radiative loss (synchrotron and IC) time for 
electrons in the bow-shock region. The advection and diffusion times are 
shown for $R_{\rm c} = 10^{10}$ (thin line) and $10^{11}$~cm
(thick line). This figure corresponds to
the case $B_{\rm bs} = 150$~G.}
\label{fig_loss_2}
\end{figure}

For electrons, the maximum energy is constrained by the escape of particles 
via diffusion
($B_{\rm bs}=1$~G) and by synchrotron radiation 
($B_{\rm bs}=150$~G), as it
is shown in Figs. \ref{fig_loss_1} and \ref{fig_loss_2}. From these 
figures, we determine that the most relevant radiative process in the
bow-shock region can be IC scattering if $B_{\rm bs}=1$~G, and synchrotron
radiation if $B_{\rm bs}=150$~G.
On the other hand, the maximum energy for protons accelerated in 
the bow shock is constrained by the Hillas criterion (Hillas 1984), 
i.e. when the proton gyroradius becomes equal to the size
of the acceleration region:

\begin{equation}
E_p^{\rm max} =x\, q\, B_{\rm bs}= 0.2\, R_{\rm c}\, q\, B_{\rm bs}\,.
\end{equation}
In Table \ref{Table_energies}, maximum
energies for electrons and protons  of different values of $B_{\rm bs}$ 
and $R_{\rm c}$ are shown.

\begin{table}[]
\begin{center}
\caption{Maximum energies achieved by acelerated particles in the bow shock. 
The values are in eV units and represent the different values 
of $R_{\rm c}$ and $B$ studied in this work.}
\label{Table_energies}
\begin{tabular}{c|cccc}
\hline 
$R_{\rm c}$ [cm] & $10^{10}$ & $10^{10}$ & $10^{11}$ & $10^{11}$ \\
$B_{\rm bs}$ [G] & $1$ & $150$ & $1$ & $150$ \\ 
\hline 
$E_e^{\rm max}$ [eV] & $1.5\times 10^{11}$ & $8\times 10^{11}$ & 
$1.5\times10^{12}$ & $8\times 10^{11}$  \\  
$E_p^{\rm max}$ [eV]& $6\times 10^{11}$ & $9\times 10^{13}$ & 
$6\times10^{12}$ & $9\times 10^{14}$  \\   
\hline
\end{tabular}
\end{center}
\end{table}

\section{Production of gamma-rays and lower energy 
radiation}
\label{gamma}

We calculated the spectral energy distribution (SED)
of the emission produced by the most relevant non-thermal radiative 
processes. In the bow-shock region, we considered synchrotron and IC 
radiation, and in the clump, we also considered 
relativistic bremsstrahlung and  $pp$.

\subsection{Distribution of relativistic particles}

We assume an injected population of relativistic particles 
(electrons and protons) in the bow-shock region
that follows a power-law energy distribution of the form
\begin{equation} 
Q_{e,p}(E_{e,p}) = K_{e,p}\,E_{e,p}^{-\Gamma}\,
\exp(-E_{e,p}/E_{e,p}^{\rm max})\;.
\end{equation}
The index $\Gamma$ is fixed to 2, 
typical of linear diffusive shock acceleration, and we add an exponential
cut-off at high energies. The normalization constant $K_{e,p}$ 
is determined by assuming that $25\,\%$ of the 
jet luminosity incident on the bow shock, i.e. 
$\sim (\sigma_{\rm c}/\sigma_{\rm j})L_{\rm kin}$ (where 
$\sigma_{\rm c}= \pi R_{\rm c}^2$ is the clump effective cross section
\footnote{We neglect here the region where the bow shock becomes 
strongly oblicuous, and focus on where this shock is the strongest,
i.e. right in front of the clump.}), 
is converted into power of the accelerated particles.
We note that the predicted fluxes 
scale linearly with the adopted non-thermal fraction.

To estimate the particle energy distribution $N_{e,p}$, we solve
the kinetic equation in the one-zone model approximation for the bow-shock
region (e.g. Ginzburg \& Syrovatskii 1964)

\begin{equation} 
\frac{\partial N_{e,p}}{\partial t} = \frac{\partial}{\partial E_{e,p}}
(P_{e,p} N_{e,p}) - 
\frac{N_{e,p}}{\tau_{\rm esc}} + Q_{e,p}\,,
\end{equation}
where $t$ is the time and $P_{e,p} = -\partial E_{e,p}/\partial t$ is the 
energy loss rate for particles.

The relativistic leptons reach the steady state well before the 
shock has crossed the clump. As shown in
Figs. \ref{fig_loss_1} and \ref{fig_loss_2}, 
the most energetic electrons can diffuse to the
clump ($B_{\rm bs} = 1$~G) or lose their energy
inside the bow-shock region by synchrotron and IC radiation 
($B_{\rm bs} = 150$~G). However,
particles downstream with low energies escape
advected in the shocked material of the jet before cooling radiatively, 
producing a break in the energy spectrum of particles.
By equating the advection and synchrotron loss times, we can 
estimate the break energy in the case with $B_{\rm bs} = 150$~G, which 
results in $E_{\rm b}=3\times 10^9$ and $E_{\rm b}=3\times 10^8$~eV,
for $R_{\rm c} = 10^{10}$ and $10^{11}$~cm, respectively.
For the case of $B_{\rm bs} = 1$~G, the break energy is determined by the
advection and diffusion times, given $E_{\rm b} = 0.3 E_e^{\rm max}$.
The electrons with energies $E_e > E_{\rm b}$ can reach
the clump and radiate inside it.
The energy distribution of these electrons is
\begin{equation}
N_e^{\rm c}(E_e) \sim Q_e^{\rm c}(E_e)\; t_{\rm cool}\,,
\end{equation} 
where $Q_e^{\rm c}$ is the part of $Q_e$ that corresponds to 
$0.3\,E_e^{\rm max}<E_e<E_e^{\rm max}$, and 
$t_{\rm cool} = 1/(t_{\rm syn}^{-1} + t_{\rm IC}^{-1} + t_{\rm Brem}^{-1} + 
t_{\rm ion}^{-1})$.
The ionization cooling term, $t_{\rm ion}\sim 2\times10^{18}\,(E_e/n)$~s, 
comes from the high densities in the clump.

As noted in Sect. \ref{cooling}, relativistic protons do not suffer
significant $pp$ losses in the bow-shock region ($t_{\rm diff} \ll t_{pp}$).
These protons can also reach the clump if they are not
advected by the shocked 
material of the jet. By assuming that $t_{\rm diff} < t_{\rm adv}$,
the minimum energy necessary to reach the clump is
\begin{equation} 
E_p^{\rm min} = 0.025 E_p^{\rm max}\,, 
\end{equation}
and the  maximum energies of these protons are determined by the 
Hillas criterion and shown in Table \ref{Table_energies}.
On the other hand, to confine relativistic protons in the clump,  
the magnetic field must be
$> 10^3$~G, much stronger than the expected clump magnetic 
field\footnote{Assuming equipartition, 
magnetic fields $\sim 100$~G would be expected.}, i.e., the protons 
will cross the entire clump in a time 
$\sim R_{\rm c}/c < t_{pp}$. We note that $R_{\rm c}$ should be the 
shocked clump width if the clump has already been affected by the shock.
The distribution of relativistic protons in the clump is

\begin{equation} 
N_p(E_p) = \frac{R_{\rm c}}{c}\,Q_p(E_p)\,.
\end{equation}
With the steady distributions of relativistic electrons in the
bow shock, $N_e(E_e)$, and of both electrons and protons in the clump, 
$N_e^{\rm c}(E_e)$ and $N_p(E_p)$, respectively,
we can calculate the SEDs of the radiation produced by these particles.

\subsection{Non-thermal emission from the bow shock}
\label{jet_nonthermal}

Since the energy density of the synchrotron emission
is lower than that of the magnetic and stellar fields, synchrotron
self-Compton processes will be neglected in our calculations.   

Synchrotron radiation is computed using the standard formulae given in
Blumenthal \& Gould (1970). Assuming that the distribution of 
relativistic electrons in
the bow-shock region is isotropic and moves with a
non-relativistic advection speed, we calculate the synchrotron SED as
\begin{equation}
\epsilon L_{\rm synch}(\epsilon) = 
\epsilon \int_{E_e^{\rm min}}^{E_e^{\rm max}} N_e(E_e)\;
P(E_e, \epsilon)\,dE_e\,,
\end{equation}
where 
\begin{equation}
P(E_e, \epsilon) = 10^{34} B_{\rm bs} 
(\epsilon/E_{\rm c})^{(1/3)} \exp(-\epsilon/E_{\rm c}) \,
\rm{s^{-1}}
\end{equation}
is the power function. The magnitude 
$E_{\rm c}(E_e)= 5.1\times10^{-8}\,B\,E_e^2$~erg is the 
characteristic energy of the produced photons. 

For a thermal distribution of target photons, we can 
estimate the IC emission from
\begin{equation}
\epsilon L_{\rm IC}(\epsilon) = u_{\rm ph}\,c \;\epsilon 
\int_{E_e^{\rm min}}^{E_e^{\rm max}} 
N_e(E_e) \frac{d\sigma(x_{\rm IC}, \epsilon_0, E_e)}{dE_e}dE_e\,,
\end{equation}
where $d\sigma(x_{\rm IC}, \epsilon_0, E_e)/dE_e$ is a parametrized 
differential cross section for an isotropic target photon field, 
valid in both Th and KN regimes (Blumenthal \& Gould 1970). This
magnitude depends on the adimensional parameter
\begin{equation} 
x_{\rm IC} = \frac{\epsilon}{4\epsilon_0\left(\frac{E_e}{m_ec^2}\right)^2
\left(1-\frac{\epsilon}{E_e}\right)}\,,
\end{equation}
for $m_e^2c^4/(4E_e^2) < x_{\rm IC} \le 1$, where
$\epsilon_0$ is the energy of the target photons.

\begin{figure}
\includegraphics[angle=0, width=0.45\textwidth]{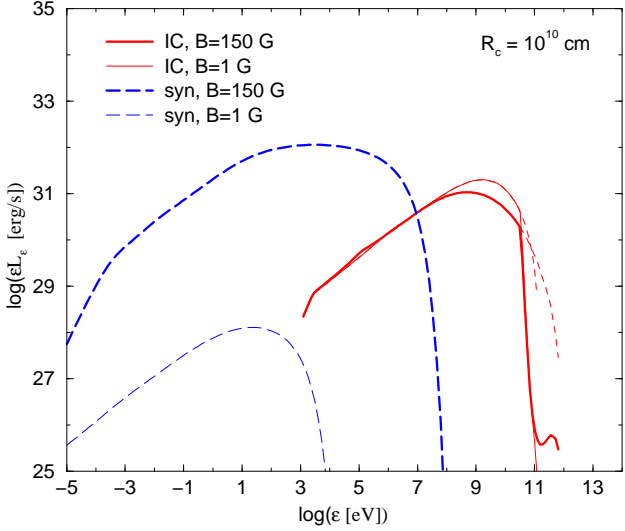}
\caption{Computed  synchrotron (dotted line) and IC (dashed line) SEDs 
for the emission produced in the bow-shock region for the 
case of $R_{\rm c} = 10^{10}$~cm with $B_{\rm bs}=1$ (thin line) and 150~G 
(thick line). The production and the $\gamma-\gamma$ absorbed  curves
at very high energies are shown.}
\label{Synch-IC_1}
\end{figure}

\begin{figure}
\includegraphics[angle=0, width=0.45\textwidth]{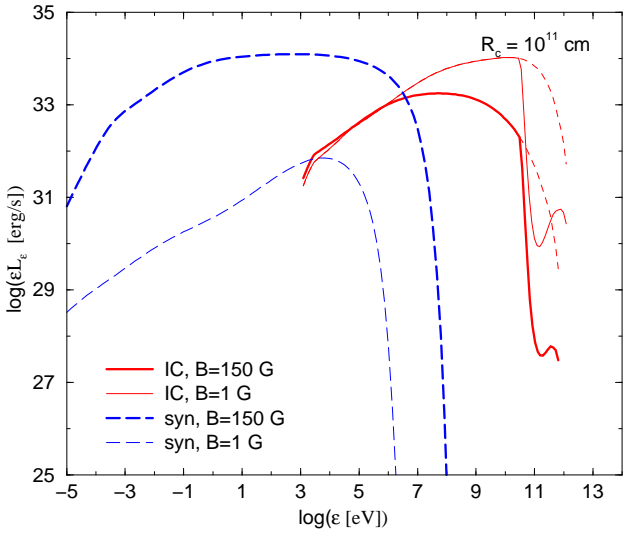}
\caption{The same as in Fig. \ref{Synch-IC_1}, but for the case 
$R_{\rm c} = 10^{11}$~cm.}\label{Synch-IC_2}
\end{figure}

We calculated $\epsilon L_{\rm synch}(\epsilon)$ and 
$\epsilon L_{\rm IC}(\epsilon)$ 
for different values of $B_{\rm bs}$ and $R_{\rm c}$, and the results
are presented in Figs. \ref{Synch-IC_1} and \ref{Synch-IC_2}.
As seen in these figures, the synchrotron  component is
more luminous than the IC one in the cases of $B_{\rm bs} = 150$~G,
reaching bolometric luminosities of 
$L_{\rm synch} \sim 10^{33}$ and $2\times10^{35}$~erg~s$^{-1}$ 
for $R_{\rm c} = 10^{10}$ and $10^{11}$~cm, respectively. 
In contrast, 
for $B_{\rm bs} = 1$~G, the dominant radiative process is IC scattering, which 
leads to bolometric luminosities of
$L_{\rm IC}\sim 2\times10^{32}$~erg~s$^{-1}$ and 
$\sim 10^{35}$~erg~s$^{-1}$ for
$R_{\rm c} = 10^{10}$ and $10^{11}$~cm, respectively.
The maximum energies achieved by photons can be as high
as $E_{\rm ph}^{\rm max} \sim 1$~TeV. We
note the spectral break in the synchrotron and IC emission for the case 
$B_{\rm bs} = 150$~G.
This feature
is produced by the advection of particles out from the 
bow-shock region.
In addition, the impact of KN losses hardening the electron spectrum
can also be seen in the spectral shape of the synchrotron and IC
emission for $R_{\rm c} = 10^{11}$~cm and $B_{\rm bs} = 1$~G.    

We accounted for $\gamma-\gamma$ absorption in the stellar field of the
massive star. Since we did not focus on a particular
binary/observer system geometry, we assumed an isotropic target
photon field (and hence also neglected angular effects in the IC calculations). 
As seen in Figs. \ref{Synch-IC_1} and \ref{Synch-IC_2},  
the attenuation in flux can be of several orders of magnitude at energies
of hundreds of GeV.
In only some cases of specific geometries in the $\gamma-\gamma$
interaction, can the attenuation be very much reduced 
(see e.g. Khangulyan, Aharonian \& Bosch-Ramon 2008). 

In this work, we have focused on the high-energy emission. 
Radio emission is also produced by electrons that leave the bow shock well
before they lose most of their energy. These particles may radiate
in the radio band further down the jet and their treatment is beyond the 
context of this study. Because of this, we have not  
considered the effect of synchrotron self-absorption in the radio
spectrum, and we do not make predictions in this energy range. 

\subsection{Non-thermal emission from the clump}
\label{clump_nonthermal}

The most energetic particles accelerated in the bow-shock region 
can penetrate the clump. In the case of protons, uncooled particles with
$E_p > 0.025 E_p^{\rm max}$ can diffuse up to the clump and radiate only a 
part of their energy from there. 
On the other hand, for $B_{\rm bs}=1\,{\rm G}$ ($\ll B_{\rm equipartition}$), 
electrons with $E_e > 0.3 E_e^{\rm max}$ will radiate their energy in the clump.

The relativistic protons that diffuse up to the clump collide with  
cold protons there, generating both neutral ($\pi^0$) and charged 
($\pi^{\pm}$) 
pions that decay to $\gamma$-rays and leptons, respectively.

\begin{figure*}
\centering 
\includegraphics[angle=0,width=0.45\textwidth]{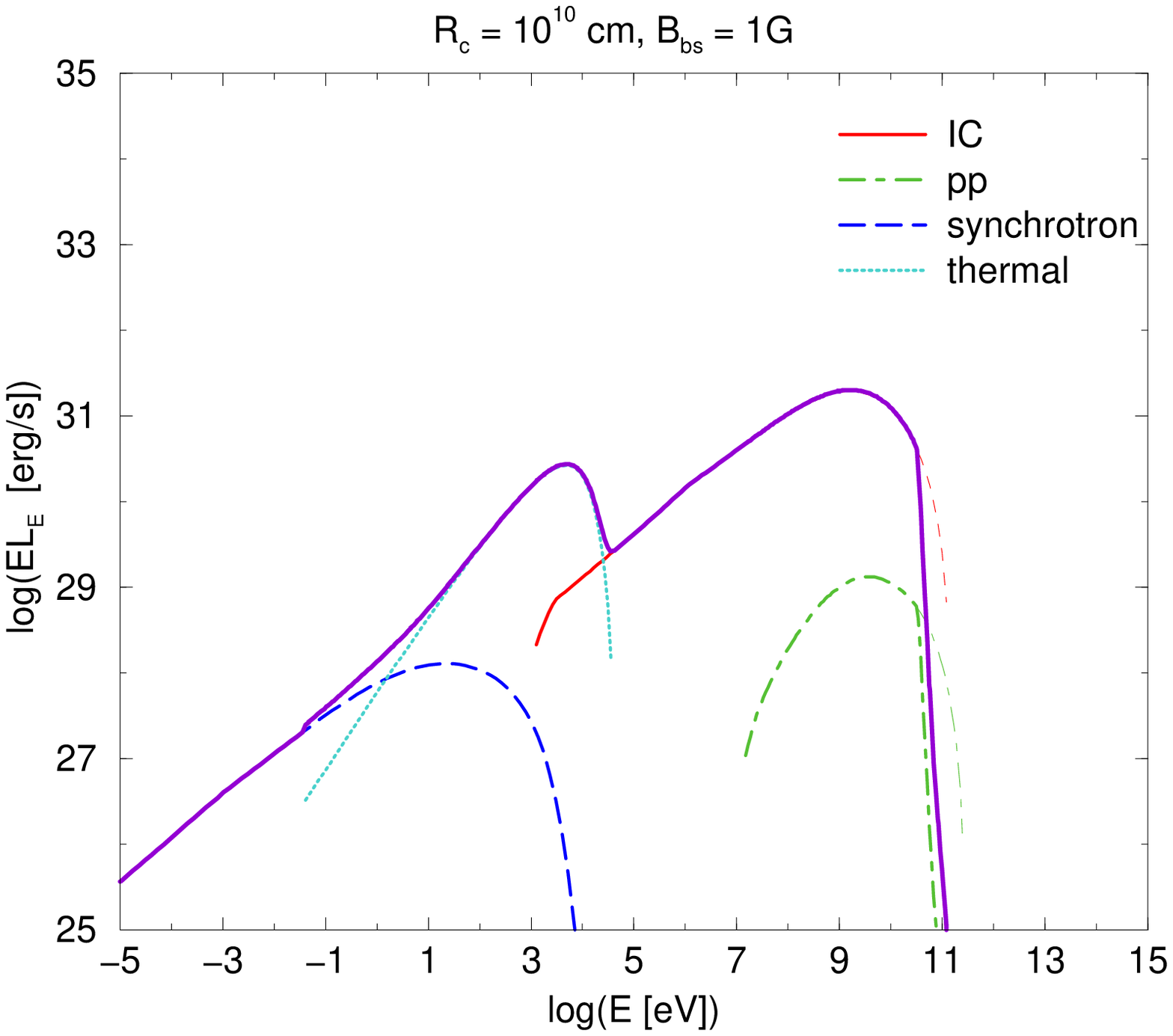}\qquad
\includegraphics[angle=0,width=0.45\textwidth]{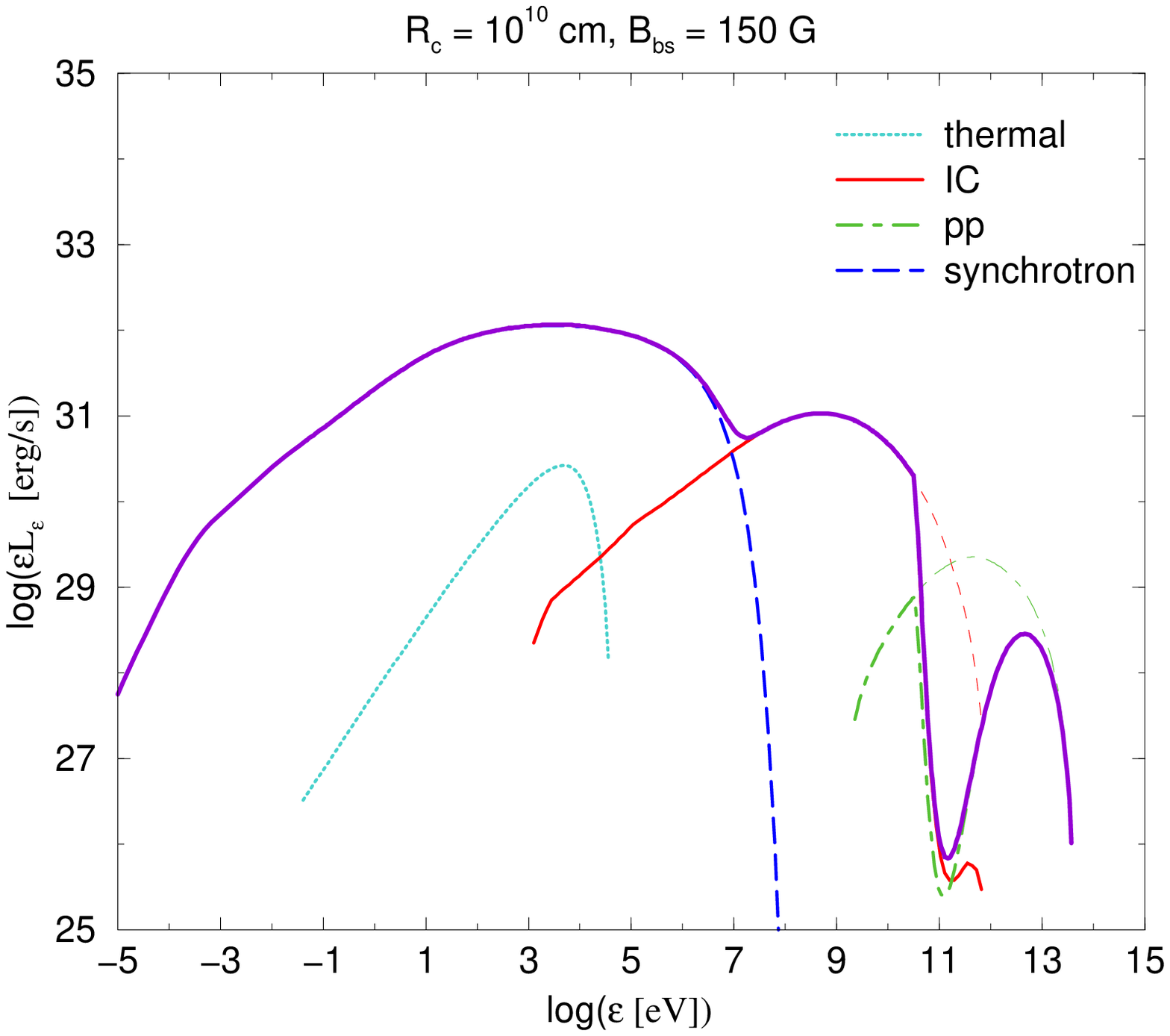}\\[10pt]
\includegraphics[angle=0, width=0.45\textwidth]{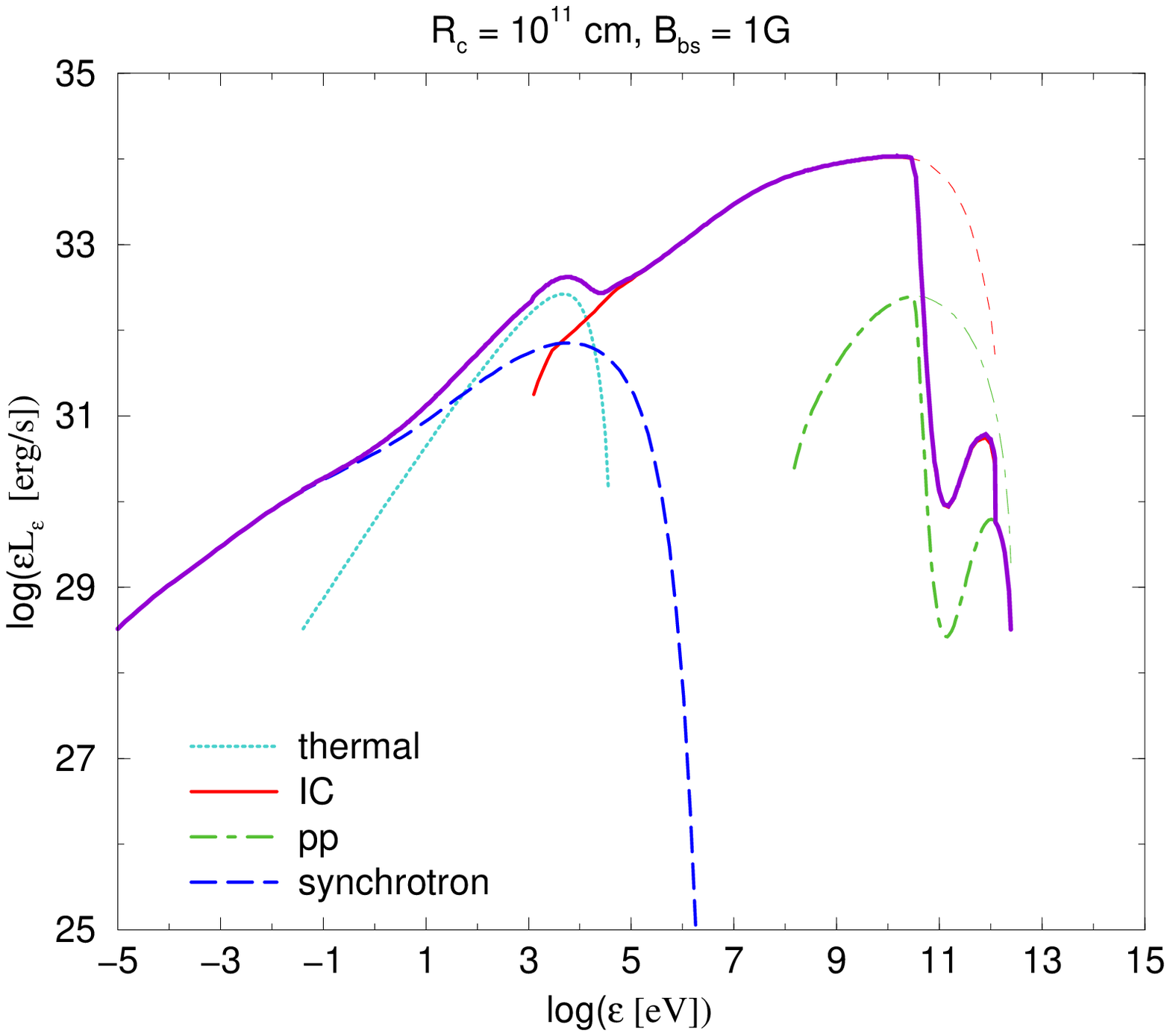}\qquad
\includegraphics[angle=0, width=0.45\textwidth]{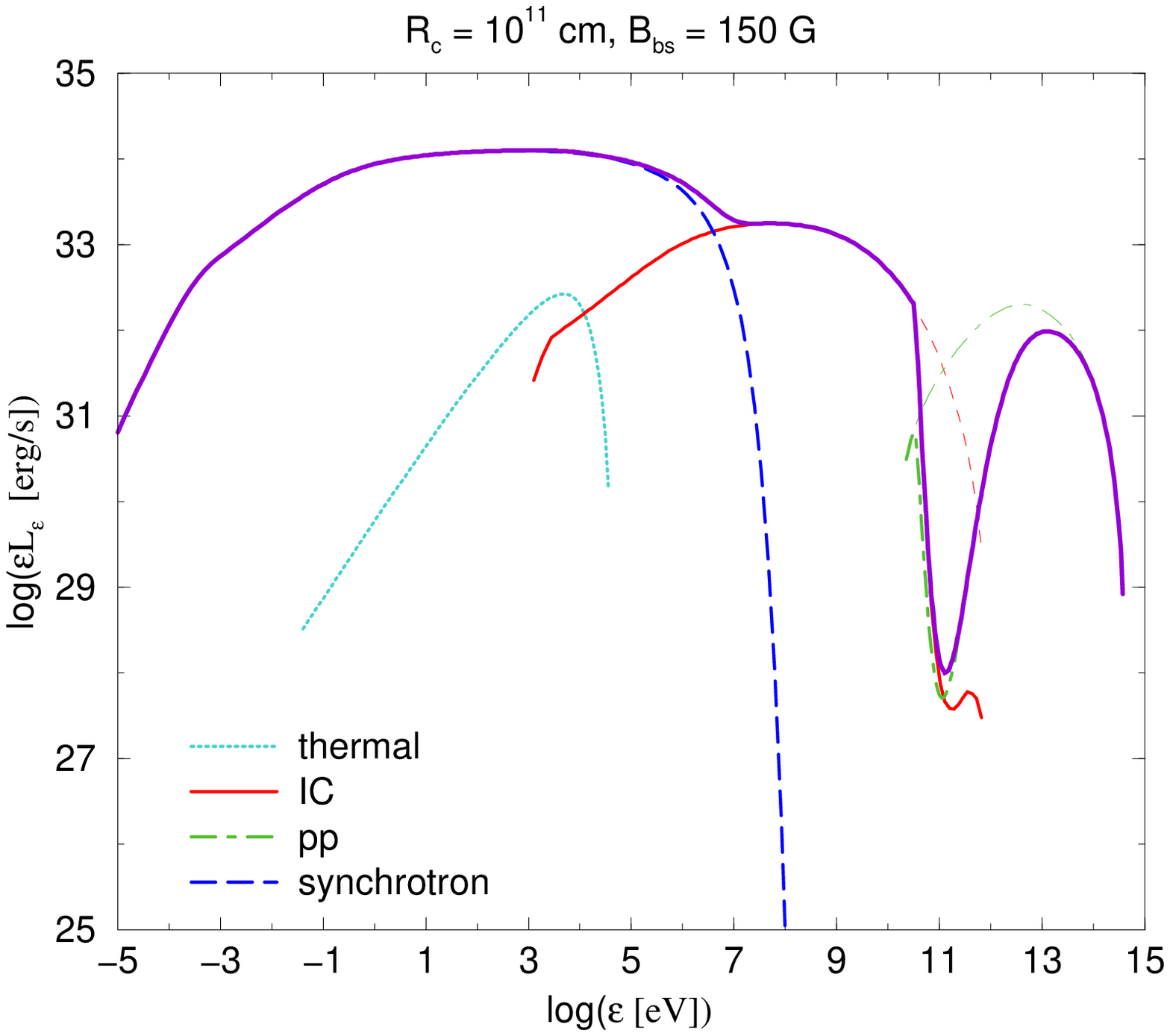}\\
\caption{SEDs for different values of $B_{\rm bs}$ and $R_{\rm c}$;
the curves of both absorbed and unabsorbed (thin lines) IC and $pp$ radiation 
are shown.}
\label{SEDs}
\end{figure*}

The emissivity of $\pi^0$ is given by

\begin{equation}
q_{\pi}(E_{\pi}) = c\; n_{\rm c} \int_{E_{p}^{\rm min}}^{E_{p}^{\rm max}}
\sigma_{pp}(E_p) N_p(E_p) dE_p\,, 
\end{equation}  
where $\sigma_{pp}(E_p)$ is the cross section of $pp$ interactions estimated
by Kelner et al. (2006). With knowledge of the emissivity of pions, 
the specific luminosity
of the photons produced by $\pi^0$-decay is   

\begin{equation}
\epsilon L_{\epsilon} = 2 \epsilon\int_{E_{\pi}^{\rm min}}^{E_{\pi}^{\rm max}}  
\frac{q_{\pi}(E_{\pi})}{\sqrt{E_{\pi}^2 - m_{\pi}^2c^4}} dE_{\pi}\,,
\end{equation}
where $E_{\pi}^{\rm min} = \epsilon + m_{\pi}^2c^4/(4\epsilon)$ and
$E_{\pi}^{\rm max} \sim E_p^{\rm max}/6$.
The estimated emission is shown in Fig. \ref{SEDs}, 
along with the contributions of the other processes.
The emission reaches luminosities of as high as 
$L_{pp} \sim 10^{32}$~erg~s$^{-1}$ ($R_{\rm c}=10^{11}$~cm; Fig. \ref{SEDs}).
For the adopted  values of $B_{\rm bs}$ and $E_p^{\rm max}$, the 
$\gamma-\gamma$ absorption reduces the final fluxes and can lead to very 
different spectral shapes. 

Secondary pairs with an effective low energy cut-off at 
$\sim 0.1E_p^{\rm min}$ will be injected inside the clump by the decay 
of $\pi^{\pm}$. These pairs radiate most of their
energy inside the clump by synchrotron, IC  scattering, and relativistic
bremsstrahlung. 
In general, primary electron emission from both the bow-shock region and
the clump (see the next paragraph) is more significant than that of
this secondary component (see Bosch-Ramon, Aharonian \&
Paredes 2005, and Orellana et al. 2007 
for discussion of secondary pairs in the context of clouds and jets,
respectively).
Finally, we note that very high-energy neutrinos with luminosities 
$\sim L_{pp}$ would also be generated 
(e.g., Aharonian et al. 2006; Reynoso \& Romero 2009).    

To estimate the radiation produced by electrons accelerated in the 
bow-shock, we assume two values of the magnetic
field of the clump: $B_{\rm c} = 1$ and $100$~G, the latter
being similar to the equipartition clump magnetic field. 

\begin{figure}
\includegraphics[angle=0, width=0.45\textwidth]{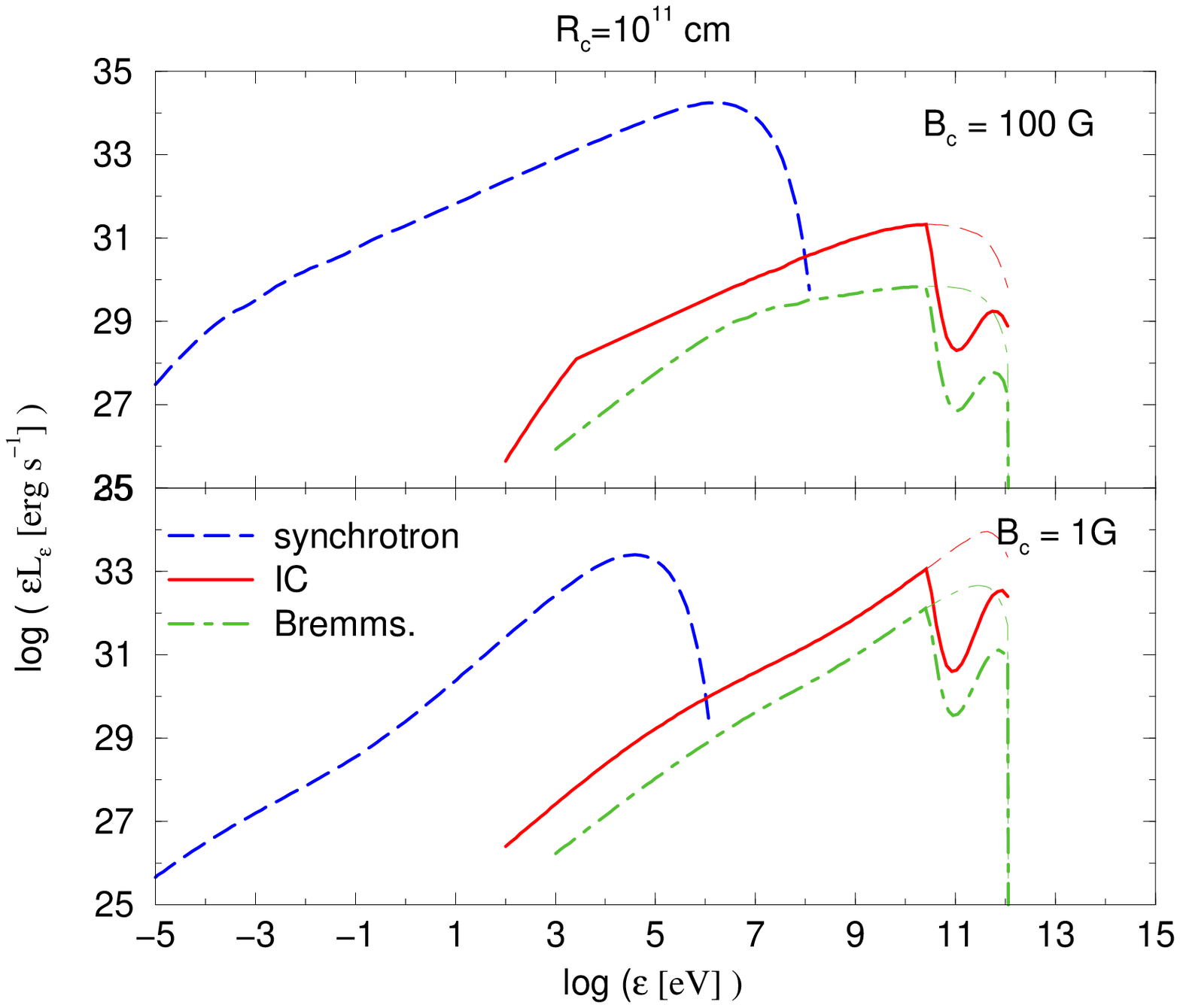}
\caption{Non-thermal leptonic emission from the clump, for the case
$R_{\rm c} = 10^{11}$~cm and for the two assumed values of $B_{\rm c}$, 1 
(bottom) and 100~G (top), is shown. The curves for both absorbed 
(thick lines) and unabsorbed (thin lines) IC radiation and 
relativistic bremsstrahlung are shown.}
\label{clump_e}
\end{figure}

The synchrotron and the IC radiation were calculated using the same 
equations given in Sect. \ref{jet_nonthermal}. Regarding the relativistic 
bremsstrahlung,
we used the standard formulae given in Blumenthal \& Gould (1970). 
As shown in Fig. \ref{clump_e}, 
synchrotron radiation is higher than IC in the case of $B_{\rm c} =
100$~G, reaching a luminosity $\sim 10^{34}$~erg~s$^{-1}$ 
($R_{\rm c} = 10^{11}$~cm) at 
$\epsilon \sim 0.1$~MeV. For $B_{\rm c} = 1$~G, on the other hand, 
the absorbed IC component reaches a similar luminosity to that of the
synchrotron one, $\sim 10^{33}$~erg~s$^{-1}$. Relativistic bremsstrahlung is
negligible in all cases. 
The SEDs for $R_{\rm c} = 10^{10}$~cm are similar to those
shown in Fig. \ref{clump_e}, but the luminosities are 
$\propto \sigma_{\rm c}$ and therefore two orders of magnitude less. 
We note that the non-thermal emission from the clump is similar to, 
but of slightly lower intensity than the bow shock one, being the 
spectra harder. 
We note that the radio emission produced by leptons in the clump is strongly
suppressed by ionization losses and the high low-energy cut-off of
the injected electrons.
For clarity, we have not plotted the leptonic emission from the clump together
with the other components, but they can be compared using Figs~\ref{SEDs} and 
\ref{clump_e}.

\section{Discussion and summary}
\label{disc}

We explored the main physical processes and the nature of the
radiation produced by the interaction between the jet of a HMMQ and a
clump of the wind of the companion star. The penetration of a clump in the
jet produces two shocks. One shock quickly reaches a steady
state in the jet, forming a bow-like shock. The other shock 
propagates through the clump driven by pressure equilibrium in the
jet/clump contact discontinuity.  The bow shock is adiabatic and fast,
and particles can be accelerated to very high energies, 
leptons efficiently emitting synchrotron and IC radiation. Otherwise,
the shock in the clump is slow and radiative. It is a non-efficient
accelerator but a thermal emitter as well as a good target for
proton-proton collisions. If electrons from the bow-shock region 
enter the clump, they could also radiate energy efficiently there 
via synchrotron and IC processes. 
The SEDs of the mentioned radiation
components have been computed in the context of a microquasar with
parameters similar to those of Cygnus X-1, showing that in some cases
the interaction between one clump and the jet can yield significant
amounts of radiation. The entire SEDs for different cases of clump size
and magnetic field are shown in Figs.~\ref{SEDs} and \ref{clump_e}.

As noted above, given the limitations of the one-zone approximation,
the synchrotron emission has been computed in the optically thin case,
neglecting the impact of synchrotron self-absorption. Nevertheless, we
note that the radio fluxes can be quite high, despite probably being
self-absorbed right at the bow-shock region. Radio emission 
produced further down the jet by
electrons accelerated in the bow shock may contribute significantly to
the radio band, i.e., jet-clump interactions should have associated
radio flares.

At X-rays, the emission is produced by synchrotron (bow shock and clump) 
and thermal radiation (clump). In the cases with $B_{\rm bs}\sim 1$~G, the
thermal emission, reaching
luminosities of $L_{\rm th}\sim 10^{32}$~erg~s$^{-1}$, 
is of higher intensity than that of synchrotron 
from the bow-shock region, but not from the clump if $R_{\rm c}=10^{11}$~cm.
Synchrotron emission from the bow-shock region
dominates in the X-rays band for $B_{\rm bs}=150$~G and
$L_{\rm synch} \sim 10^{35}$~erg~s$^{-1}$. In a source such as
Cygnus~X-1, these levels of X-ray emission would be overcome by the
accretion disk radiation. Nevertheless, in the case of
fainter jet sources at
X-rays, such as LS~5039 and LS I +61 303 (Bosch-Ramon et al. 2007, 
Paredes et al. 2007), the synchrotron X-rays
produced during the jet-clump interaction may be detectable, and even
the thermal component may be detectable on the top of a non-thermal
component in certain conditions (large clumps of relatively low
densities).

Inverse Compton scattering in the bow-shock region and the 
clump produces $\gamma$-rays up to 
very high energies that dominate the radiation output in cases of
relatively weak magnetic fields (e.g., $B_{\rm bs}=1$~G). In our
calculations, the highest luminosity achieved is $L_{\rm IC}\sim
10^{35}$~erg~s$^{-1}$ for $R_{\rm c}=10^{11}$~cm, although
$\gamma-\gamma$ absorption can reduce the emission substantially above
100~GeV. Proton-proton collisions in the clump can also produce 
$\gamma$-rays at energies that may be as high as $\sim
10^{14}$~eV ($B_{\rm bs}=150$~G). The maximum luminosity obtained by
$pp$ is nevertheless quite modest, $L_{pp} \sim 10^{32}$~erg~s$^{-1}$
for $R_{\rm c}=10^{11}$~cm, although denser and/or bigger clumps, and
more powerful jets may yield detectable amounts of photons outside 
the $\gamma-\gamma$ absorption range (0.1-10~TeV). 
We recall that specific geometries of the binary/observer
system plus a high-energy emitter far from the compact object may 
render the attenuation of $\gamma$-rays much smaller 
(e.g., Khangulyan et al. 2008). 

As a consequence of the characteristics of the interaction, the expected
emission is transient. The flare duration is related to the permanence
of the clump inside the jet dependent on RT
and KH instabilities, which can destroy the clump. Since the clump 
is accelerated inside the jet, it may be disrupted after several dynamical
timescales. If the clump were not destroyed, it would eventually leave
the jet as a thin and moderately hot slab of plasma, since the shock in 
the clump
is radiative. Given the typical dynamical and crossing timescales,
the entire event may have a duration of between a few minutes and hours.

The flares of jet-clump interactions would have associated lower
(synchrotron, thermal emission) and higher energy components (IC,
$pp$ emission), which should not be correlated with the 
accretion disk activity. 
The total level of emission,
the importance of the different components, and the duration of the
flares, can provide information about the jet power and size, clump size and
density, and magnetic fields in the interaction regions 
(Romero et al. 2007). Therefore,
besides the jet itself, the clump properties can be probed by
observations at high and very high energies (and probably also at
radio frequencies) of transient activity in HMMQs, which are a new tool for
studying the winds of massive stars.

Depending on the wind filling factor (or clump density) and the 
clump size/number -$N_{\rm c}$- (Owocki \& Cohen 2006), the 
flares produced by interactions of clumps with jets in HMMQs 
could be a sporadic phenomenon, for a low number of clumps, 
or may appear as a modulated steady activity, for a high number
of clumps (see Owocki et al. 2009). For the parameters adopted
here, the former case corresponds to $R_{\rm c}=10^{11}$~cm
and the latter to $R_{\rm c}=10^{10}$~cm. Nevertheless, we note that
the jet may be disrupted in those cases when too many clumps are 
simultaneously present inside the jet. Assuming that jet disruption
takes place for $\sigma_{\rm j} < N_{\rm c}\times\sigma_{\rm c}$,
for the wind and jet properties adopted in this work, the jet could be
destroyed if $R_{\rm c}\la 10^{10}$~cm. 
However, more detailed calculations of the dynamics of the jet-clump 
interaction are required to clarify this issue.  

\begin{acknowledgements}
%The authors thank an anonymous referee for his/her constructive comments.
The authors thank Stan Owocki and Dmitry Khangulyan for many insightful 
discussions on clumps, jets, and hydrodynamics.
A.T.A. thanks the Max Planck Institut f$\rm\ddot{u}$r Kernphysik for 
its suport and kind hospitality.
V.B-R. and G.E.R. acknowledge support by DGI
of MEC under grant AYA2004-07171-C02-01, as well as partial support by
the European Regional Development Fund (ERDF/FEDER).
V.B-R. gratefully acknowledges support from the Alexander von Humboldt
Foundation.

\end{acknowledgements}

{}
\end{document}